\documentclass[fleqn,twoside]{article}
\usepackage{espcrc2}
\usepackage{latexsym}
\usepackage{euscript}

\usepackage{graphicx}
\usepackage[figuresright]{rotating}

\def\simge{\mathrel{%
   \rlap{\raise 0.511ex \hbox{$>$}}{\lower 0.511ex \hbox{$\sim$}}}}   
\def\simle{\mathrel{   
   \rlap{\raise 0.511ex \hbox{$<$}}{\lower 0.511ex \hbox{$\sim$}}}}   
\def\slashchar#1{\setbox0=hbox{$#1$}           
   \dimen0=\wd0                                 
   \setbox1=\hbox{/} \dimen1=\wd1               
   \ifdim\dimen0>\dimen1                        
      \rlap{\hbox to \dimen0{\hfil/\hfil}}      
      #1                                        
   \else                                        
      \rlap{\hbox to \dimen1{\hfil$#1$\hfil}}   
      /                                         
   \fi}                                         %

\def\simge{\mathrel{%
   \rlap{\raise 0.511ex \hbox{$>$}}{\lower 0.511ex \hbox{$\sim$}}}}   
\def\simle{\mathrel{   
   \rlap{\raise 0.511ex \hbox{$<$}}{\lower 0.511ex \hbox{$\sim$}}}}   
\def\slashchar#1{\setbox0=\hbox{$#1$}           
   \dimen0=\wd0                                 
   \setbox1=\hbox{/} \dimen1=\wd1               
   \ifdim\dimen0>\dimen1                        
      \rlap{\hbox to \dimen0{\hfil/\hfil}}      
      #1                                        
   \else                                        
      \rlap{\hbox to \dimen1{\hfil$#1$\hfil}}   
      /                                         
   \fi}

\newcommand{\AmS}{{\protect\the\textfont2
  A\kern-.1667em\lower.5ex\hbox{M}\kern-.125emS}}

\newcommand{\ket}[1]{\left| {#1} \right\rangle}
\newcommand{\bra}[1]{\left\langle {#1}\right|}
\newcommand{\ba}{\begin{equation} \left\{ \begin{array}{lr}}
\newcommand{\ea}{\end{array} \right. \end{equation}}
\newcommand{\nn}{\nonumber}
\newcommand{\bea}{\begin{eqnarray}}
\newcommand{\eea}{\end{eqnarray}}
\newcommand{\be}{\begin{equation}}
\newcommand{\ee}{\end{equation}}
\newcommand{\id}{\mbox{1$\!\!$I}}
\newcommand{\Tr}{\mbox{Tr}\;} 
\newcommand{\Pj}{\mbox{I}\!\!\mbox{P}}
\newcommand{\pslash}{{\slashchar p}} 
 
\newcommand{\de}{\partial}

\newcommand{\RI}{\mbox{\scriptsize RI}}

\title{Numerical Exploration of the RI/MOM Scheme Gauge Dependence\thanks{Talk given at Lattice 2002 by N.T.}}

\author{L. Giusti\address[CERN]{ Theory Division, CERN, 
        CH-1211 Geneva 23, Switzerland}\address[CNRS]{Centre de Physique 
                Th\'eorique, CNRS Luminy, Case 907, F-13288 Marseille Cedex 9, France},
        S. Petrarca\address[ROME1]{Dipartimento di Fisica, Universit\`a di Roma ``La Sapienza'', 
        P.le A. Moro 2, I-00185 Rome, Italy}\address[INFNR1]{INFN Roma 1, 
        P.le A. Moro 2, I-00185 Rome, Italy},
        B. Taglienti\addressmark[INFNR1]
        and
        N. Tantalo\address[ROME2]{Dipartimento di Fisica, Universit\`a di Roma ``Tor Vergata'',
        V. R. Scientifica 1, I-00133 Rome, Italy}\address[INFNR2]{INFN Roma 2, 
        V. R. Scientifica 1, I-00133 Rome, Italy}}
       
\begin{document}

\begin{abstract}
The gauge dependence of some fermion
bilinear RI/MOM renormalization constants is studied
by comparing data which have been gauge-fixed in two
different realizations of the Landau gauge and  
in a generic covariant gauge.
The very good agreement between the various sets of results
and the theory indicates that the numerical uncertainty
induced by the lattice gauge-fixing 
procedure is below the statistical errors
of our data sample which is of the order of $(1 \div 1.5)\% $.
\vspace{-0.2cm}
\end{abstract}

\maketitle

\section{Introduction and Strategy}
Non-perturbative renormalization techniques 
\cite{Martinelli:1994ty,luscher_np} have become a crucial 
ingredient in lattice computations of fundamental QCD parameters and hadronic 
matrix elements. 
The RI/MOM non--perturbative renormalization proposed in Refs.~\cite{Martinelli:dq,Martinelli:1994ty}
has been successfully applied to compute renormalization constants
of composite fermion operators in many lattice regularizations \cite{Martinelli:dq,Martinelli:1994ty,Aoki:1999mr,Blum:2001sr,Giusti:2001pk}.
The method imposes renormalization conditions on conveniently defined 
amputated Green functions  computed 
non-perturbatively between off-shell quark states at large virtuality
in a fixed gauge:
\begin{equation}
Z_\Gamma \left.\bra{p} 
O_\Gamma \ket{p}\right|_{p^2=\mu^2}
= \bra{p} O_\Gamma \ket{p}_0 \;.
\end{equation}
The renormalization constants 
depend in general on the external states and therefore on the gauge in which 
the renormalization conditions have been imposed. 
For gauge-invariant operators
the coefficients needed to match in a given gauge-invariant scheme has to cancel these dependences 
up to higher orders in the continuum perturbative expansion and up to discretization errors.

The existence of both continuum and lattice Gribov
copies and the numerical noise 
that they can generate 
(for a recent review see Ref.~\cite{gf_review}) 
is an unsolved problem of the 
lattice non-perturbative gauge fixing. 
The real concern is, of course, 
the influence that these phenomena may have on the values
of the renormalization constants, when computed using non-gauge-invariant 
quantities as in the RI/MOM scheme.

Here we discuss the results of a numerical
study of the systematics induced by the 
gauge-fixing procedure on the RI/MOM determinations of 
the quark field ($Z_\psi$), 
the axial-vector ($Z_A$) and of the scalar density ($Z_S$) renormalization constants
(for more details see Ref.~\cite{Giusti:2002rn}).
We have done this by imposing the standard lattice Landau gauge 
and another realization of the
Landau condition (Landau1 gauge).
Whereas the two realizations impose the same gauge-fixing condition in the naive continuum
limit, they are affected by different Gribov ambiguities. 
By comparing the different sets of results for 
the renormalization constants, 
we have found differences which are
negligible within our statistical errors (of the order of $(1 \div 1.5)\%$).

An interesting feature of
the Landau1 gauge is that it can be generalized to
impose a generic covariant gauge on the lattice, as proposed in 
Ref.~\cite{Giusti:1999im}. By exploiting this opportunity, 
we have performed an exploratory study of the 
gauge dependence of off-shell Green functions
measured in a generic covariant gauge.

\section{Results and Discussion}
We have used a sample of $80$ SU(3) gauge configurations retrieved from the
repository at the ``Gauge Connection'' (http://qcd.nersc.gov/), which were  
generated with the Wilson gluonic action at $\beta=6.0$
and $V \times T = 16^3 \times 32$. By using the discretized 
version (see~\cite{Giusti:1999im} for details) of the 
following functionals: 
\be
F_A[G] = ||A^G||^2=
\int d^4x\, \mbox{\rm Tr}\left[A^G_{\mu}A^G_{\mu}\right]  
\\ \nn
\ee
\be
H_A [G] = 
\int d^4x\,\mbox{\rm Tr}\left[(\partial_{\mu}A^G_{\mu})^2\right]
\\ \nn
\ee
\be
H_{(A,\Lambda)}[G] = 
\int d^4x\, \mbox{\rm Tr}\left[(\partial_{\mu}A^G_{\mu}-\Lambda)^2 \right]\; \;,
\label{eq:lambdax}
\ee
we have rotated each configuration 
in the Landau gauge, in the Landau1 gauge  and 
in the covariant gauge.
The function $\Lambda$ belongs to the Lie algebra of the
group and has been generated with a gaussian distribution
corresponding to the value $\xi=8$ of the bare gauge fixing parameter. 

\begin{figure}[htb]
\vspace{-1.1cm}
\hspace{-2.8cm}\includegraphics[width=12cm]{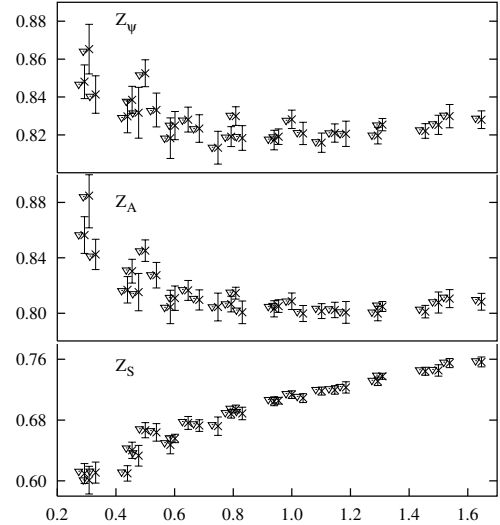}
\vspace{-1.5cm}
\caption{\small The $Z_A$ and $Z_S$ renormalization constants are
shown as a function of $(a \mu)^2$ for the Landau (crosses) and 
Landau1 (triangles) gauges.
The data have been slightly displaced in the $x$-direction and
the error bars for one set only have been shown.}\vspace{-0.5cm}
\label{fig:ssig1}
\end{figure}

\begin{figure*}[ht]
\includegraphics[width=16cm,height=6.5cm]{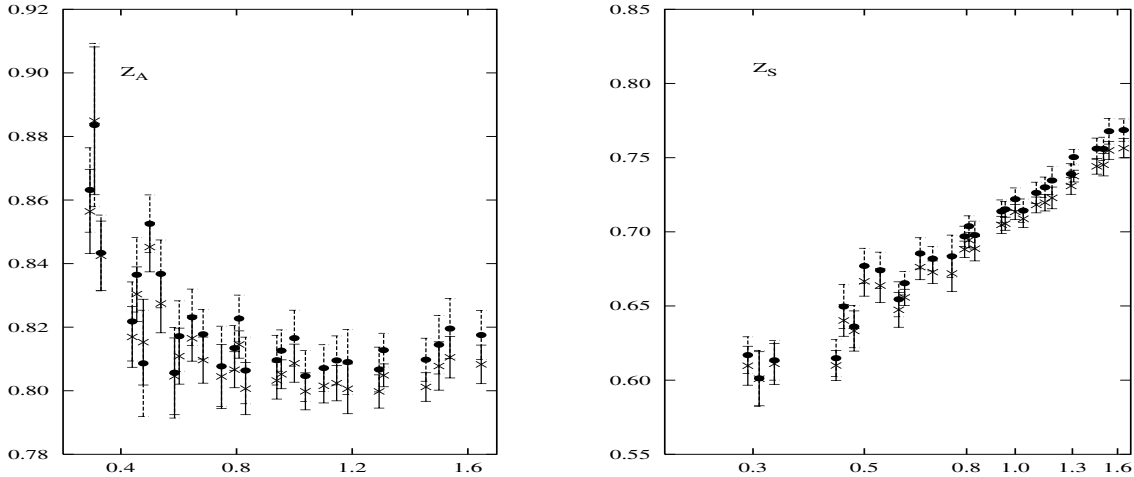}
\label{fig:ZaZs}
\vspace{-1cm}
\caption{\small The behaviour of $Z_A$ and $Z_S$  is
shown as a function of $(a \mu)^2$ for the  Landau (black dots)
and the covariant gauge (crosses) data. The residual gauge-dependence is
of the order of 1.5\%}
\end{figure*}

Once the quark propagator $\mathcal{S}(x,0)$ 
has been computed for each SU(3) configuration and gauge and Fourier-transformed, we have determined the  
inverse Euclidean bare-quark propagator $\mathcal{S}^{-1}(pa)$.
We have computed the amputated Green functions of the local quark bilinears
$O_\Gamma(x)=\bar{\psi}(x) \Gamma \psi(x)$ in the Fourier 
space (in what follows we adopt the conventions
of Ref.~\cite{Gimenez:1998ue}) 
\begin{equation}
\Lambda_\Gamma(pa)=\mathcal{S}^{-1}(pa)G_\Gamma(pa)\mathcal{S}^{-1}(pa)\; ,
\label{eq:lambda}
\end{equation}  
where
\be
G_\Gamma(p) =
\int{
d^4x d^4y \,e^{-ip(x-y)}\langle \psi(x) O_\Gamma(0) \bar{\psi}(y) \rangle} 
\ee
and $\Gamma$ is a generic Dirac matrix 
corresponding to 
$\Gamma = \left\{\mbox{A},\mbox{S} \right\} = \left\{\gamma_\mu\gamma_5,\id\right\}$.
The renormalization constant $Z_\Gamma^{\RI}(\mu a, g_0)$, which defines the 
renormalized operator $\hat {\cal O}_\Gamma^{\RI} =Z_\Gamma^{\RI} {\cal O}_\Gamma$,
is fixed by imposing in the chiral limit the renormalization condition
\begin{equation}
Z_\Gamma^{\RI}(\mu a)Z_{\psi}^{-1}(\mu a) \Tr\Pj_\Gamma\Lambda_\Gamma (pa)
|_{p^2=\mu^2}=1,
\label{eq:RI}
\end{equation}
where $\Pj_\Gamma$ is a suitable projector on the tree-level amputated
Green function and where
the RI wave-function renormalization constant $Z_\psi$ is defined as 
\begin{equation}
Z_{\psi}(\mu a)=\lim_{m\rightarrow 0}-i
\frac{1}{12} \Tr \left[\frac{\de S^{-1}(pa)}{\de \pslash }\right]_{p^2=\mu^2}\; .
\label{eq:Z_q_WI} 
\end{equation}

Even though ${\cal O}_A$ and ${\cal O}_S$ are gauge-independent
operators, their matrix elements between quark states
are gauge dependent. In Fig.~{\ref{fig:ssig1}}, the numerical
results for $Z_\psi$, $Z_A$ and $Z_S$ calculated
in the Landau and in the Landau1 gauge as a function of the square lattice momenta, are shown.
Our data are in very good agreement with the results reported 
in the literature~\cite{Gimenez:1998ue}.

The results for all correlators, corresponding to Landau and Landau1 gauges,
coincide within the statistical errors. As a consequence we can conclude that 
the data do not show any residual effect due to the presence of 
lattice  Gribov's copies in the statistical sample of configurations
generated and for the lattice used.

To show our sensitiveness to the choice of the gauge, 
we compare the results obtained 
in the Landau and in the covariant gauge with the bare 
gauge parameter fixed to $\xi=8$. 

The primary quantities measured on the lattice,  
$Z_\psi$, $\Lambda_A$ and $\Lambda_S$,
show a statistically 
significant gauge dependence (as you can see in Fig.~2 of Ref.~\cite{Giusti:2002rn}) while the renormalization constants 
$Z_A$, $Z_S$ shown in Fig.~\ref{fig:ZaZs}, obtained by computing the ratios 
as indicated in Eq.~(\ref{eq:RI}), could have at most 
a residual gauge dependence (of the order of 1.5\%)
which is not detectable within the statistical precision of our
simulation. For these quantities, the fluctuations of
the simulation hide the weak gauge 
dependence that is expected in
perturbation theory from the next-to-leading order 
terms in $\alpha_s$.

These results indicate an upper limit to the 
numerical troubles that can be expected in the
RI/MOM renormalization constants because
of the lattice gauge fixing. 

\vspace{-0.2cm}


\end{document}